\documentclass[final,namedreferences,hyperref,optionalrh]{spr-sola}

\usepackage{graphicx}
\usepackage{color}

\usepackage{amsmath}
\usepackage{amssymb}
\usepackage{bm}

\usepackage{float}

\usepackage{booktabs}
\usepackage{multirow}
\usepackage{array}
\usepackage{tabularx}

\renewcommand{\vec}[1]{{\mathbfit #1}} 

\chardef\us=`\_

\usepackage[a4paper, top=2.5cm, bottom=2.5cm, left=2.5cm, right=2.5cm]{geometry}

\begin{document}

\begin{frontmatter}
\title{Generalization of Elliptical-Cylindrical Flux Rope Models for ICME Reconstruction}

\author[addressref={aff1,aff2},corref,email={marti.masso.moreno@estudiantat.upc.edu}]{\inits{M.}\fnm{Martí}~\snm{Massó~Moreno}}
\author[addressref={aff2,aff3},email={cperezal@gmu.edu}]{\inits{C.A.}\fnm{Carlos~Arturo}~\snm{Pérez-Alanis}}
\author[addressref={aff2,aff4},email={manoharanp@cua.edu}]{\inits{P.K.}\fnm{P.~K.}~\snm{Manoharan}\orcid{0000-0003-4274-211X}}

\address[id=aff1]{CFIS, Universitat Politècnica de Catalunya (UPC), Barcelona, Spain}
\address[id=aff2]{Heliophysics Science Division, NASA Goddard Space Flight Center, Greenbelt, MD 20771, USA}
\address[id=aff3]{George Mason University, Fairfax, VA 22030, USA}
\address[id=aff4]{The Catholic University of America, Washington, DC 20664, USA}

\runningauthor{Massó~Moreno et al.}
\runningtitle{Generalization of Elliptical-Cylindrical Flux Rope Models for ICME Reconstruction}

\begin{abstract}
We present a generalized elliptical–cylindrical flux–rope model for interplanetary coronal mass ejections (ICMEs) that allows for a nonzero poloidal component in the internal magnetic field. A two–step reconstruction algorithm is introduced, decoupling the geometric configuration from the magnetic–field fitting to improve numerical stability and physical consistency. Applied to Parker Solar Probe data, the new radial–poloidal model preserves the global flux–rope geometry while achieving a substantially better fit to the internal magnetic field than the traditional radial model, offering a more accurate and realistic description of ICME structures.
\end{abstract}


\end{frontmatter}

\section{Introduction}
\label{S-Introduction}
Interplanetary Coronal Mass Ejections (ICMEs) are large expulsions of plasma from the solar corona that can be detected in situ by spacecraft in the interplanetary medium. Observationally, such CMEs often show an initially irregular shock front followed by a smoother, more coherent magnetic structure known as a Magnetic Flux Rope (MFR). An MFR can be broadly defined as a flux tube whose magnetic field lines twist around a single common axis \citep{2024ApJ...xxx...yyyW}. 

Flux rope observations can be interpreted using a variety of geometries and internal physical constraints. Some of the most common analytical descriptions are the linear force–free model of \citet{Lundquist50} (often referred to as the Lundquist model) and the force–free uniform twist model of \citet{1960MNRAS..120...89G}. More recently, models with more relaxed constraints or more generic geometries have been developed, for example the circular–cylindrical model \citep{2016ApJ...823...27N} and the elliptic–cylindrical model \citep{2018ApJ...861..139N}, as well as subsequent extensions and applications \citep[e.g.][]{2023ApJ...947...79N}. Nevertheless, these approaches still enforced an internally purely radially defined geometry by imposing that the radial component of the current density vanish. 

In this work, we develop a generalization that allows for a non-zero radial component of the current density and present a new reconstruction method, alternative to the global algorithms used thus far. This method ensures convergence of the solutions and enables a comprehensive analysis of all possible geometric configurations corresponding to the spacecraft’s trajectory through the MFR.

\section{Generalization of the Elliptical Cylindrical model}

\subsection{Elliptical-Cylindrical Coordinate System}

Considering the above hypotheses, we will attempt to generalize the possible internal magnetic‐field configurations while maintaining the local cylindrical–elliptical geometry. The cylindrical–elliptical coordinates and their contravariant basis vectors are defined as follows:
\begin{equation}
    x = a\, r \, \cos\varphi, \qquad  y = y, \qquad  z = \delta \, a\, r \, \sin\varphi.
\end{equation}
\begin{equation}
    \vec{\varepsilon}_r = a \, ( \cos\varphi,\; 0 ,\; \delta \, \sin\varphi), \quad
    \vec{\varepsilon}_y = ( 0,\; 1,\; 0), \quad
    \vec{\varepsilon}_{\varphi} = a\, r \, ( - \, \sin\varphi,\; 0,\; \delta \, \cos\varphi).
\end{equation}

Here, $a$ is the semi‐major axis of the elliptical section, $r$ is the radial coordinate, $\delta$ is the distortion parameter of the section, and $\varphi$ is the poloidal angle within the section.

The metric tensor in this geometry is

\begin{equation}
g =
\begin{pmatrix}
g_{rr} & 0 & g_{r\varphi} \\
0 & g_{yy} & 0 \\
g_{\varphi r} & 0 & g_{\varphi\varphi}
\end{pmatrix}
=
\begin{pmatrix}
a^2 (\cos^2\varphi + \delta^2 \sin^2\varphi) & 0 & a^2 r \sin\varphi \cos\varphi (\delta^2 - 1) \\
0 & 1 & 0 \\
a^2 r \sin\varphi \cos\varphi (\delta^2 - 1) & 0 & a^2 r^2 (\sin^2\varphi + \delta^2 \cos^2\varphi)
\end{pmatrix}.
    \label{metricTensor}
\end{equation}

Its determinant is $\det(g) = a^4 r^2 \delta^2$, and the magnitude of a physical vector (such as the magnetic field) in this metric is given by:

\begin{equation}
    \lvert \mathbf{B} \rvert
    = \sqrt{\,g_{rr} (B^r)^2 + g_{yy} (B^y)^2 + g_{\varphi\varphi} (B^\varphi)^2 + 2\,g_{r\varphi}\,B^r B^\varphi\,}
    \label{eq_modB_in_Cyl}
\end{equation}

\subsection{Magnetic Field Requirements}

For the model to be valid, it must satisfy specific conditions governing both the magnetic field
configuration and the associated current density field. To ensure the magnetic field is physically realizable within the specified geometry, the primary
condition is that its divergence must be zero, as mandated by Gauss’s law for magnetic fields:

\begin{equation}
    \nabla \cdot \mathbf{B} = \partial_r (r B^r) + \partial_\varphi (r B^\varphi) + \partial_y (r B^y) = 0
    \label{eq_divB1}
\end{equation}

We assume, consistent with prior models (\cite{2018ApJ...861..139N}, \cite{2024ApJ...xxx...yyyW}), that no physical properties vary along the axial direction of the flux rope (i.e., $\partial_y = 0$). Additionally, each magnetic field component is expressed as a product of separable functions of the radial coordinate $r$ and azimuthal angle $\varphi$. Generally, we write:

\begin{equation}
    \begin{cases}
        B^r(r, \varphi) = f_1(r) g_1(\varphi) + K_1 \\
        B^y(r, \varphi) = f_2(r) g_2(\varphi) + K_2 \\
        B^\varphi(r, \varphi) = f_3(r) g_3(\varphi) + K_3
    \end{cases}
    \label{eq_generalBfield}
\end{equation}

Then, the divergence equation can be reduced to the following partial-differential equation:

\begin{equation}
    g_1(\varphi)\,\partial_r\!\bigl(r f_1(r)\bigr) \;+\; r f_3(r)\,\partial_\varphi\!\bigl(g_3(\varphi)\bigr) \;=\; 0
\end{equation}

And when solving it, we get the following relations:

\begin{equation}
    g_3(\varphi) = C_1 \;-\; k \int g_1(\varphi)\,d\varphi  \qquad \qquad     f_1(r) = \frac{k}{r}\int r f_3(r)\,dr \;+\; \frac{C_2}{r}
\end{equation}

Where $C_1$ and $C_2$ are integration constants, and $k$ is the constant used for solving the PDE. 

The next step now will be ensuring continuity at the boundary of the Flux Rope. The condition to be satisfied will be: 

\begin{equation}
    B^r(R, \varphi) = 0 = \left[ \frac{k}{R} \int_0^R r\,f_3(r)\,dr + \frac{C_2}{R}\right]\!g_1(\varphi) + K_1
\end{equation}

Since the term multiplying \(g_1(\varphi)\) is nonzero, the only solution is for both \(g_1(\varphi)\) and \(K_1\) to be identically zero. This implies that the poloidal component now takes the form:

\begin{equation}
    B^\varphi(r, \varphi) = f_3(r)\,C_1 + K_3 \;\equiv\; f_3(r)
\end{equation}

where, for simplicity, we have absorbed the constants \(C_1\) and \(K_3\) into the function \(f_3(r)\).

This leaves us with the following form for the magnetic-field components:

\begin{equation}
\begin{cases}
    B^r = 0\\
    B^y(r, \varphi) = f_2(r)\,g_2(\varphi) + K_2\\
    B^\varphi(r) = f_3(r)
\end{cases}
\label{magneticFieldEqs}
\end{equation}

\subsection{Current Density Derivation}

For the case of the current density vector \(\vec{j}\) within the flux rope, it must be ensured that no singularities or discontinuities arise. This requires that all components be smooth throughout the volume and that the limits exist as \(r \to 0\).  The current density components are obtained via the generalized Ampère’s law in the elliptical‐cylindrical metric:
\begin{equation}
    \nabla \times \vec{B} = \frac{\varepsilon^{ijk}}{\sqrt{\det(g)}}\,\partial_j\bigl(g_{kl} B^l\bigr) = \mu_0 \,\vec{j},
    \label{eq:LeviCivita}
\end{equation}

where \(\varepsilon^{ijk}\) is the Levi‐Civita tensor in the coordinate basis \(\{i,j,k\}=\{r,y,\varphi\}\), \(g_{kl}\) is the metric tensor, and \(\sqrt{|g|}=a^2 r \delta\) is the square root of its determinant. Here, \(g_{kl}B^l\) denotes the covariant components of the magnetic field (recalling the conversion: \(B_i=g_{il}B^l\)). At the end, we obtain these components:
\begin{align}
    j^r &= \frac{1}{a^2 \delta r \mu_0} (-\partial_\varphi B^y) \\
    j^y &= \frac{1}{a^2 \delta r \mu_0} \left[ \partial_\varphi (g_{r\varphi} B^\varphi) - \partial_r (g_{\varphi\varphi} B^\varphi) \right] \\
    j^\varphi &= \frac{1}{a^2 \delta r \mu_0} \partial_r B^y
    \label{eq_currents}
\end{align}

And substituting the magnetic field from \ref{magneticFieldEqs} and the metric tensor components from \ref{metricTensor}, we obtain: 

\begin{equation}
  \begin{gathered}
    j^r(r,\varphi)
      = \frac{-f_2(r)\,g'_2(\varphi)}{a^2\,\delta\,r\,\mu_0}\\[6pt]
    j^y(r,\varphi)
      = \frac{1}{\delta\,\mu_0}\Bigl[
          f_3(r)\,(\delta^2 - 1)\,(2\cos^2\varphi - 1)
          - (\sin^2\varphi + \delta^2\cos^2\varphi)\,(2f_3(r) + r\,f_3'(r))
        \Bigr]\\[6pt]
    j^{\varphi}(r,\varphi)
      = \frac{f'_2(r)\,g_2(\varphi)}{a^2\,\delta\,r\,\mu_0}
  \end{gathered}%
\end{equation}

\subsection{Current Density Requirements}

To ensure the current density field is physically valid, its components must have a finite limit throughout the domain, particularly at the center (\( r = 0 \)), where singularities could arise. For this limit to exist, considering that trigonometric functions are not constant, each component must satisfy:

\begin{equation}
    \lim_{r \to 0} j^i(r, \varphi) = 0, \quad i = r, y, \varphi
\end{equation}

For this to happen, all trigonometric functions must be multiplied by a radial function that vanishes as \(r \to 0^+\).

Two types of solutions will be considered for this purpose: polynomial and exponential. Let's now examine the conditions each must satisfy:

\begin{itemize}
  \item \textbf{From the \(j^r\) component} we extract that:
    \begin{itemize}
      \item \emph{Polynomial case:} $\quad f_2(r)\propto r^{n_2},\quad n_2>1$
      
      \item \emph{Exponential case:} $\quad f_2(r)\propto e^{-1/r^\alpha},\quad \alpha>0$
    \end{itemize}

  \item \textbf{From the \(j^y\) component} we see that:
    \[
      f_3(r)\propto r^{n_3},\quad n_3>0,
      \quad\text{or}\quad
      f_3(r)\propto e^{-1/r^\alpha},\quad \alpha>0,
    \]
    which ensures \(j^y\to0\) as \(r\to0\).

  \item \textbf{Finally, from the \(j^{\varphi}\) component} it is required that the minimum order of \(f_2(r)\) be strictly greater than 2. Since \(n_2>2\) for \(j^\varphi\) is more restrictive than \(n_2>1\) for \(j^r\), we adopt $ n_2>2$.
\end{itemize}




With this, we now know all the necessary restrictions to ensure the existence of the magnetic field in the elliptic–cylindrical structure under consideration. Any function satisfying these conditions will be physically valid.

\section{Iterative Reconstruction Algorithm}

To reconstruct the internal magnetic field of the Flux Rope and define both its orientation and the spacecraft’s trajectory within it, we iterate over all possible geometric configurations of the structure. For each configuration, we determine which one yields the best fit between the model and the experimental data.

In practice, the spacecraft crosses the Flux Rope and provides data in a standard coordinate system (GSE or RTN). However, the magnetic models described above are defined in an internal elliptical-cylindrical coordinate system, meaning we must first transform the observational data into these internal coordinates for a meaningful fit.

This transformation involves first converting from GSE coordinates to the local Cartesian frame of the cylinder via Euler rotation matrices, followed by a transformation from Cartesian to elliptical-cylindrical coordinates. These rotations are described by the Euler angles $\theta_x$, $\theta_y$, and $\theta_z$.

\begin{figure}[H] 
  \centering

  \subfigure{%
    \includegraphics[width=0.6\linewidth]{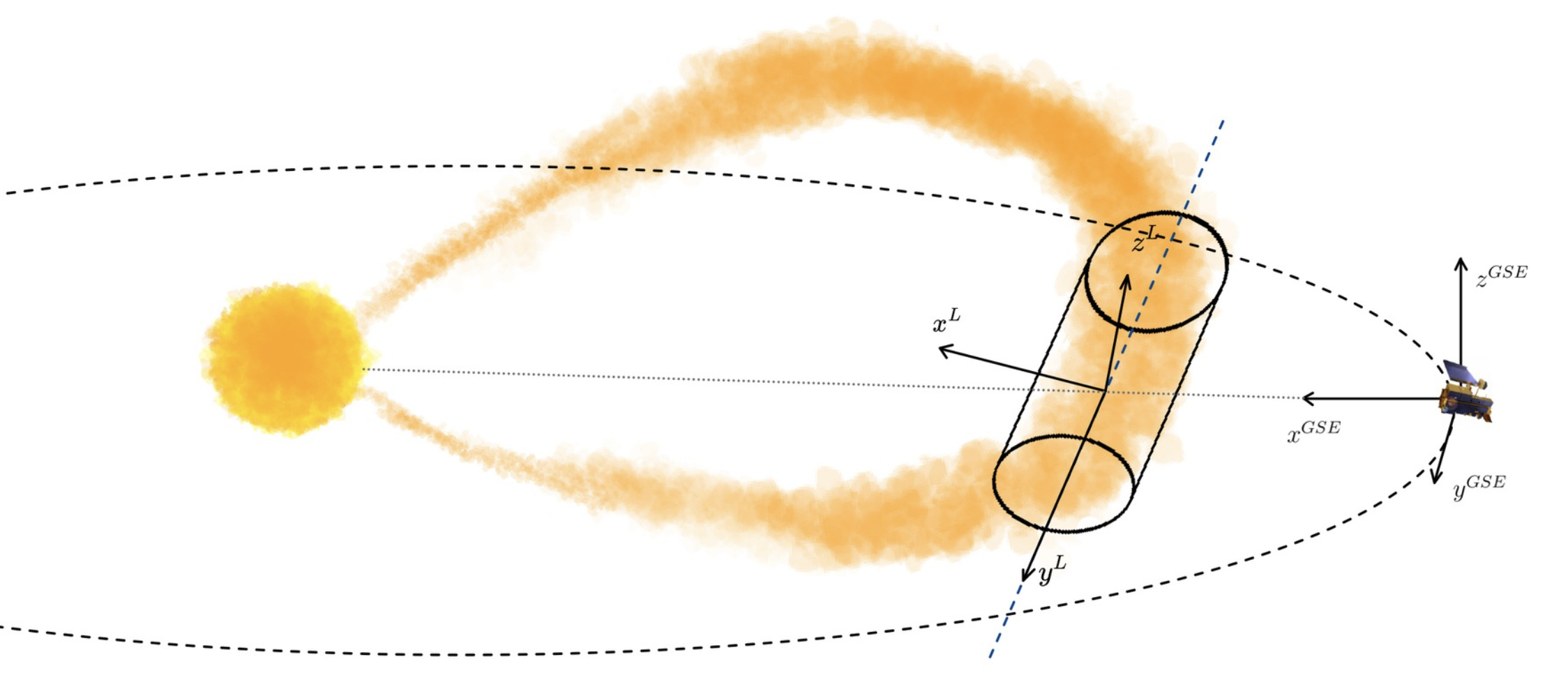}%
    \label{fig:fr-illustration}%
  }\endsubfigure
  \hfill
  \subfigure{%
    \includegraphics[width=0.3\linewidth]{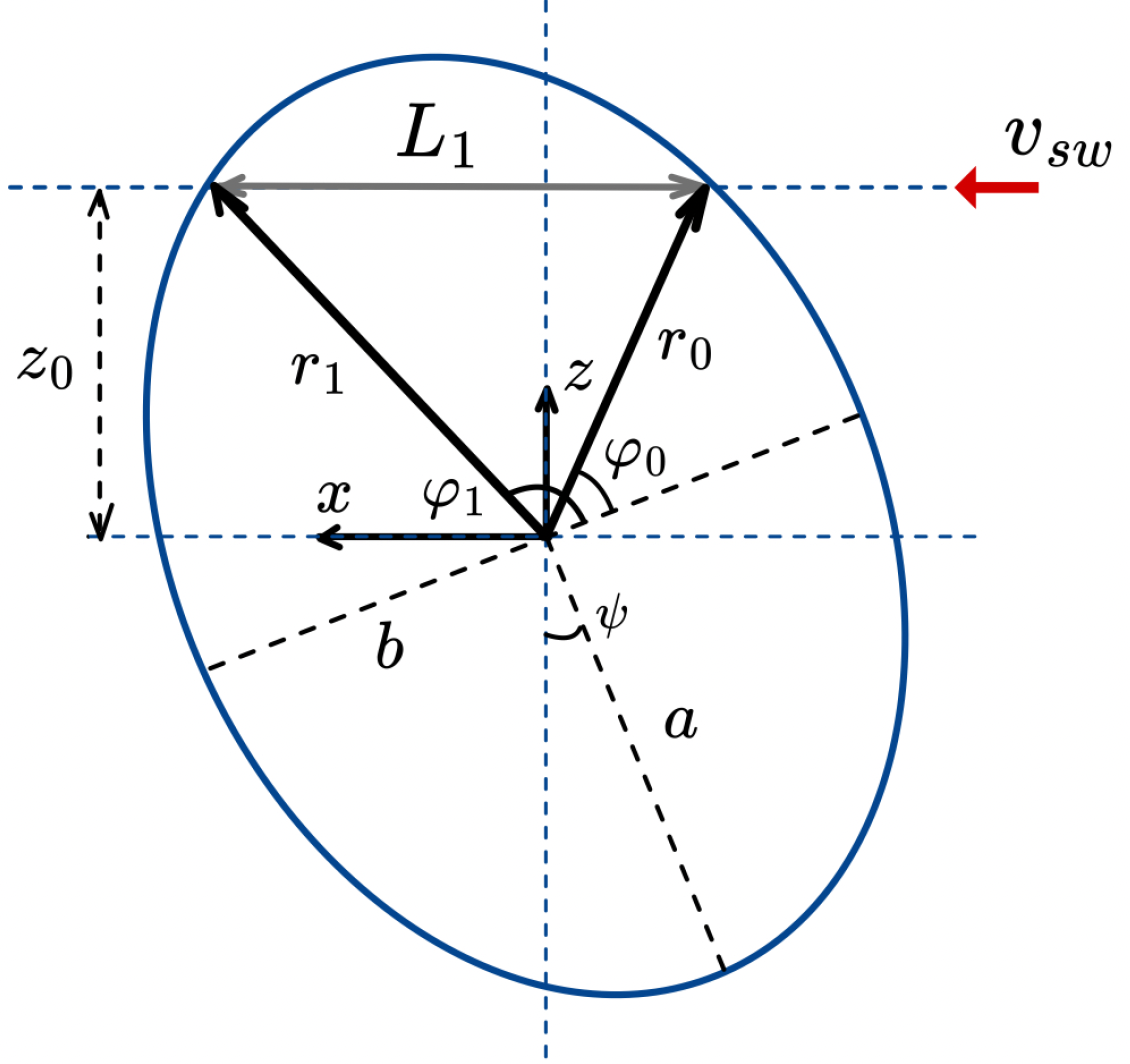}%
    \label{fig:fr-cross-section}%
  }\endsubfigure

  \caption{(a) Local Approximation of a Flux Rope inside a CME. (b) Cross section parametrization.}
  \label{fig:fr-both}
\end{figure}

The relationship between these systems is obtained by projecting the cylindrical components onto the Cartesian vectors:
\begin{equation}
    B^x_L = \hat{x} \cdot \vec{B}_{\mathrm{cyl}}, \quad B^y_L = \hat{y} \cdot \vec{B}_{\mathrm{cyl}}, \quad B^z_L = \hat{z} \cdot \vec{B}_{\mathrm{cyl}}
    \label{eq:local_cyl}
\end{equation}
\begin{equation}
\begin{pmatrix}
B_x \\
B_y \\
B_z
\end{pmatrix}_{\!L} =
\begin{pmatrix}
a\cos\varphi_{sat} & 0 & -a\,r_{sat}\sin\varphi_{sat} \\
0 & 1 & 0 \\
a\,\delta\sin\varphi_{sat} & 0 & \delta\,a\,r_{sat}\cos\varphi_{sat}
\end{pmatrix}
\begin{pmatrix}
B^r \\
B^y \\
B^{\varphi}
\end{pmatrix}_{\mathrm{Cyl}}
\label{eq:cyl_to_local}
\end{equation}

Here, $r_{sat}$ and $\varphi_{sat}$ define the spacecraft’s trajectory through the Flux Rope, $a$ is the major axis of the elliptical section, and $\delta$ is the distortion parameter of the ellipse.

\subsection{Analytical Solution to the Trajectory for a Given Configuration}

A given geometric configuration, defined by $\theta_x$, $\theta_y$, $\theta_z$, $\bar{z}_0$, and $\delta$, is resolved by computing the intersection of the elliptical cylinder with a plane containing the $x$-axis. Points along the trajectory are projected onto the FR’s cross-section, assuming axial invariance of physical properties. Using the total rotation matrix $R(\theta_x, \theta_y,\theta_z) = R_z R_y R_x$, the rotated cylinder is parameterized as:

\begin{equation}
\begin{pmatrix} x \\ y \\ z \end{pmatrix} = R \begin{pmatrix} a \cos\varphi \\ y_c \\ a \delta \sin\varphi \end{pmatrix}
\end{equation}

Where $y_c$ is the local axial component of the cylinder. This expression expands to a full trigonometric parameterization involving the rotation matrix coefficients. We simplify the notation by writing:

\begin{equation}
\begin{pmatrix} x \\ y \\ z \end{pmatrix} =
\begin{pmatrix} a A \cos\varphi + B y_c + a C \sin\varphi \\
                a D \cos\varphi + E y_c + a F \sin\varphi \\
                a G \cos\varphi + H y_c + a I \sin\varphi \end{pmatrix}
\end{equation}

Imposing boundary conditions, we determine $a$, $\varphi_0$, $\varphi_1$, $y_{c,0}$, $y_{c,1}$, and $z_0$ analytically using:

\begin{itemize}
  \item Known spacecraft internal trajectory total length $L_1 = v_{sw}\Delta t = x(\varphi_1, y_{c,1}) - x(\varphi_0, y_{c,0})$.
  \item $y = 0$ plane intersection condition to find $y_c(\varphi)$.
  \item Constant $z_0$ trajectory to find $\varphi_0$, $\varphi_1$: $cos\varphi = \frac{KG\pm I \sqrt{G^2 + I^2 - K^2}}{G^2 + I^2}$, where $K = \frac{z_0 - H y_c}{a}$.
  \item Maximum height method to compute $z_{max}$, by applying $\frac{\partial z}{\partial \varphi} = 0 \Rightarrow \text{tan}\varphi_{z_{max}} = I/G$, and finally, $z_0 = \bar{z}_0 z_{max}$.
\end{itemize}

By applying all these boundary conditions, we are ready to obtain a full description of the spacecraft trajectory through the cross-section of the Flux Rope. Now, for a generic $r$ and $\varphi$ inside this cross section, we can apply the sine and cosine theorems: 

\begin{equation}
    r^2 = a^2 + r_0^2 - 2 x_s r_0 \cos(\varphi_0), \hspace{2cm} \frac{x_s}{\sin(\varphi - \varphi_0)} = \frac{r}{\sin(\varphi_0)}
\end{equation}

And obtain the parametrization of the full trajectory contained in the cross plane $y = 0$:

\begin{equation}
    r^2 \left[ \left( \frac{\sin(\varphi - \varphi_0)}{\sin\varphi_0} \right)^2 - 1 \right] + 2 r_0 \cos(\varphi - \varphi_0) \cdot r = r_0^2 \quad \Rightarrow \quad r = r(\varphi)
\end{equation}

Finally, to obtain the final trajectory within a generic cross-section of the cylinder, we project the computed elliptical section (at $y=0$) onto the plane orthogonal to the cylinder axis. This axis is given by:
\begin{equation}
    \vec{n}_{\mathrm{axis}} = R \begin{pmatrix} 0 \\ 1 \\ 0 \end{pmatrix}
\end{equation}

The projected point \( \vec{p}_{\text{proj}} \) is computed using:
\begin{equation}
    \vec{p}_{\text{proj}} = \vec{p} - \frac{\vec{n} \cdot \vec{p}}{\|\vec{n}\|^2} \vec{n}
\end{equation}

With $\vec{p} = (x(\varphi), 0, z(\varphi))$, this gives the projection of each point of the section onto the transversal plane. This will be applied to $r(\varphi)$ to obtain the parametrization of the trajectory we need.

\subsection{Fitting of the Experimental Data}

\begin{figure}[h]
    \centering
    \includegraphics[width=0.99\linewidth]{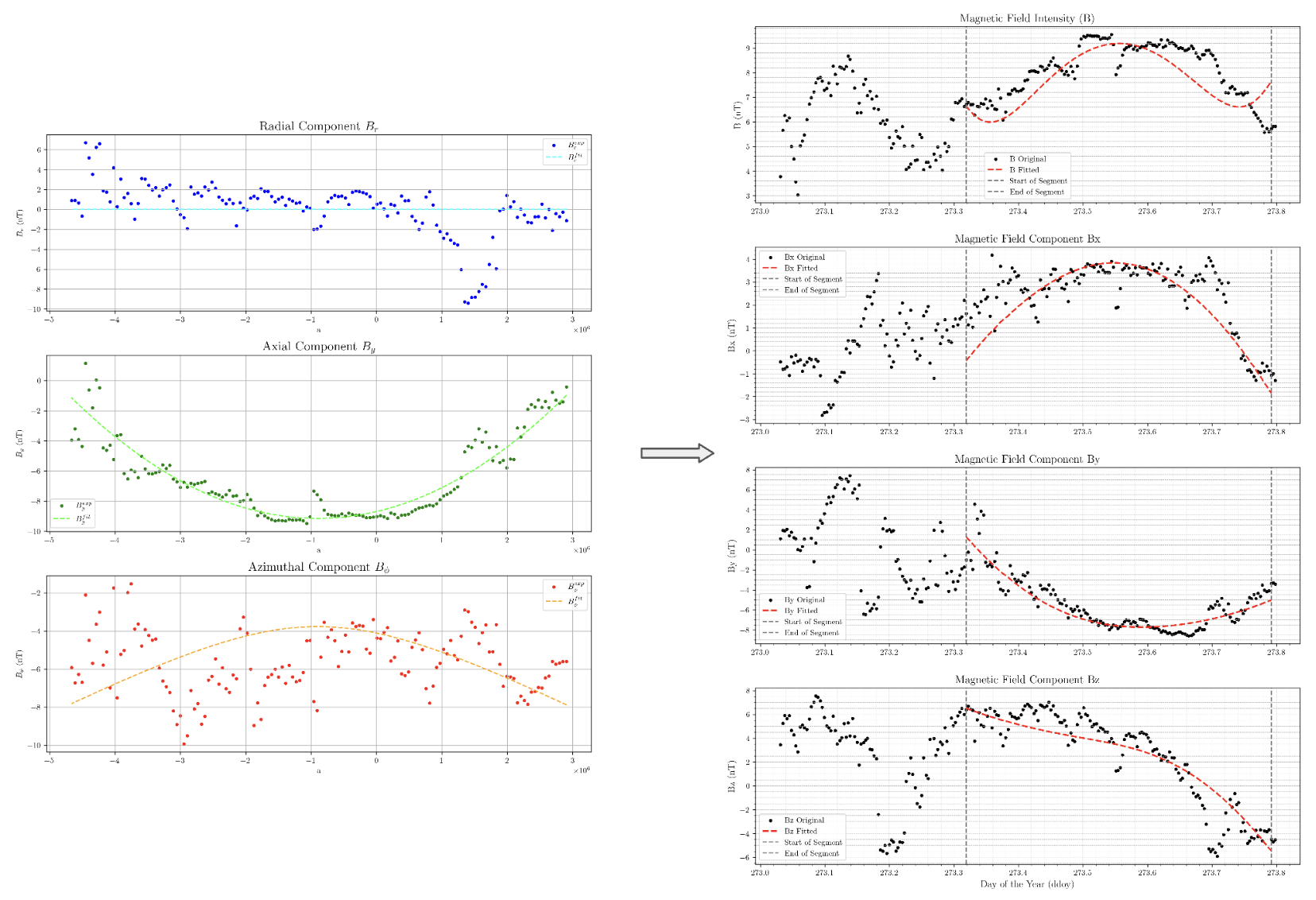}
    \caption{Fit of the values in the local cylindrical coordinate system and their subsequent representation in the GSE system.}
    \label{changing_coordinates_of_data}
\end{figure}

Once the internal trajectory has been fully determined, we transform the experimental data into the flux–rope elliptical–cylindrical frame $\bigl(B^{r}_{\exp},\,B^{y}_{\exp},\,B^{\varphi}_{\exp}\bigr)$ using relations \ref{eq:local_cyl} and \ref{eq:cyl_to_local}. After that, we fit the model to the data by minimizing a quadratic error via Gauss–Newton and Levenberg–Marquardt algorithms \cite{2018ApJ...861..139N}, as illustrated in Fig.\ \ref{changing_coordinates_of_data}.

The ultimate goal is not just a good local fit, but one that transforms back into the original GSE frame with physical consistency. We evaluate model quality by maximizing the average coefficient of determination:

\begin{equation}
R^2 = 1 - \frac{\sum_{i=1}^n (y_i - \hat{y}_i)^2}{\sum_{i=1}^n (y_i - \bar{y})^2},
\qquad R^2_{\rm avg} = \frac{1}{4}(R^2_B + R^2_{B_x} + R^2_{B_y} + R^2_{B_z})
\end{equation}

By testing all configurations and maximizing $R^2_{\text{avg}}$, we identify the most likely physical scenario, thus obtaining the flux rope orientation and internal magnetic structure. 

\subsection{Methodological Innovation Compared to Previous Approaches}

The algorithm presented in this work introduces a significant methodological advancement over previous approaches used for Flux Rope reconstruction, such as that developed by Nieves-Chinchilla et al. In particular, prior methods attempted to simultaneously fit both geometric and physical parameters in a single global optimization process. This strategy forced the numerical solver to navigate a high-dimensional, degenerate parameter space where strong correlations between orientation, impact parameter, and internal magnetic properties often led to convergence toward non-physical local minima.

In contrast, our methodology introduces a sequential decomposition of the problem:

\begin{itemize}
  \item \textbf{First, geometric configuration:} The orientation and trajectory of the spacecraft are determined analytically for every configuration in a systematic scan of the configuration space. This includes a full geometric parametrization of the rotated elliptical cylinder and its intersection with the spacecraft path, projected onto the transverse plane.
  \item \textbf{Second, physical adjustment:} Once the orientation is fixed and the data has been transformed to the internal coordinate system, the physical parameters of the magnetic field are fitted using robust local optimization, decoupled from geometric distortions.
\end{itemize}

This two-step process eliminates degeneracy between variables, enhances numerical stability, and allows for automation over large datasets. This is, to the best of our knowledge, the first implementation of a fully parameterized, two-stage, and physically consistent fitting framework for elliptical-cylindrical Flux Rope structures.

\section{Application to In Situ data}

We will do a comparison of the results obtained using two different magnetic field models, applied in our reconstruction algorithm. We will compare the previous radial model from \cite{2018ApJ...861..139N} with the proposed radial-poloidal model, for the event detected by PSP on September 16th, 2023: 


\[
\begin{array}{c @{\qquad} c}
\textbf{Radial model} 
&
\textbf{Radial–poloidal model}
\\[1ex]
\begin{cases}
B^r = 0,\\
B^y(r) = A_1 + C_1\,r^2,\\
B^\varphi(r) = D_1\,r,
\end{cases}
&
\begin{cases}
B^r = 0,\\
B^y(r,\varphi) = A\,r^3\,(B + C\cos\varphi + D\sin\varphi) + E,\\
B^\varphi(r) = F\,r + G\,r^2 + H\,r^3,
\end{cases}
\end{array}
\]

\begin{figure}[h]
    \centering
    \includegraphics[width=0.99\linewidth]{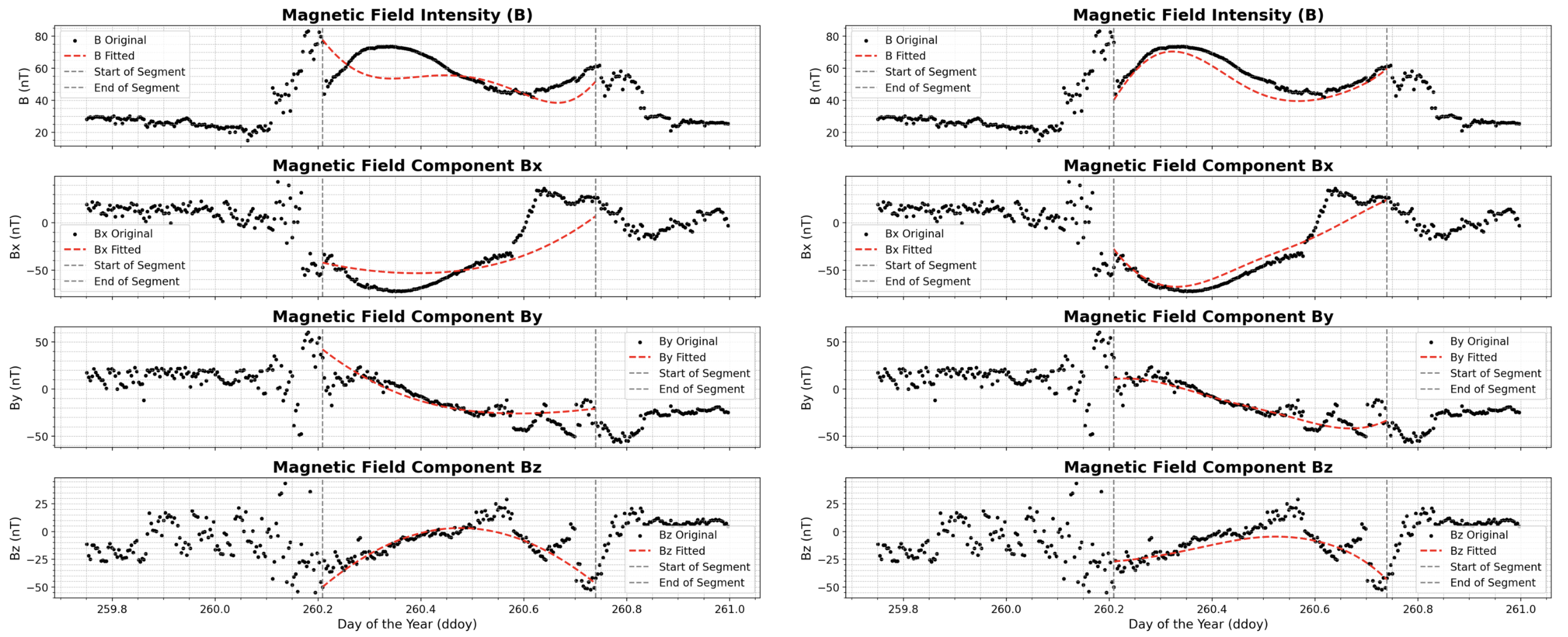}
    \caption{(a) EC geometry with radial model. (b) EC geometry with radial-poloidal model.}
    \label{fig:enter-label}
\end{figure}

\begin{figure}[H]
    \centering
    \includegraphics[width=0.99\linewidth]{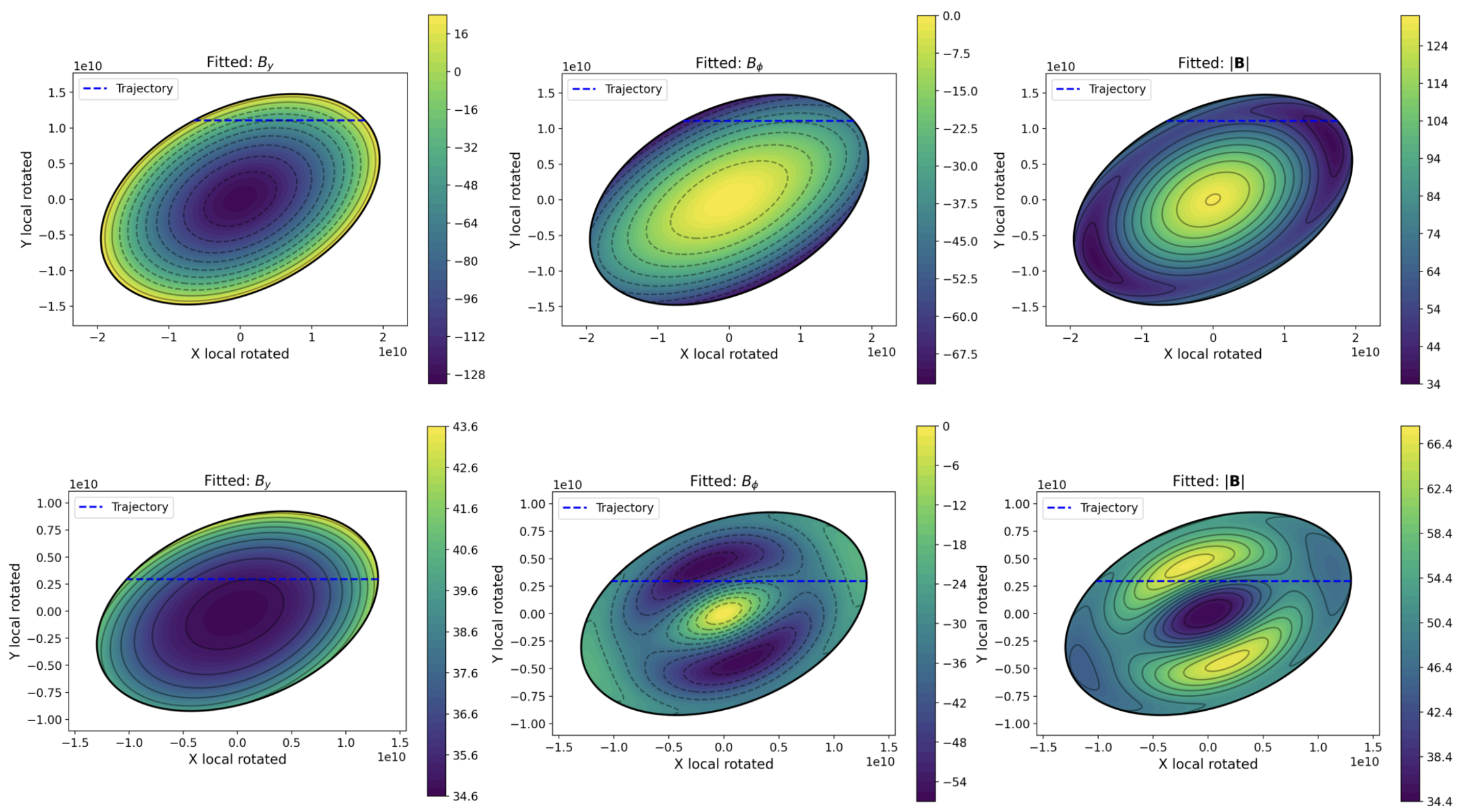}
    \caption{Comparison of the cross–sectional magnetic fields. Top panels correspond to the Radial model, whereas bottom panels show the Radial–Poloidal model.}
    \label{fig:Bfield_comparison}
\end{figure}

In Figure \ref{fig:Bfield_comparison}, we present the magnetic fields obtained from each model, along with their cross–sectional representations and the main characteristics of the resulting Flux Rope.

The analytical expressions derived for each model are given by:
\[
\begin{array}{c @{\qquad} c}
\textbf{Radial model} 
&
\textbf{Radial–Poloidal model}
\\[1ex]
\begin{cases}
B^r = 0,\\
B^y(r) = -130.42 + 3.54\cdot 10^{-19}\,r^2,\\
B^\varphi(r) = -3.46\cdot 10^{-9}\,r,
\end{cases}
&
\begin{cases}
B^r = 0,\\[4pt]
B_y(r, \phi) = 5.53 \cdot 10^{-31} r^3 \bigl( \sin(\phi - 9.54 \cdot 10^{-32}) 
\\ \qquad\qquad + \cos(\phi - 9.54 \cdot 10^{-32}) + 5.0 \bigr) + 34.72,\\[4pt]
B_\phi(r) = -2.80 \cdot 10^{-28} r^3 + 7.75 \cdot 10^{-18} r^2 - 5.61 \cdot 10^{-8} r
\end{cases}
\end{array}
\]

Table \ref{tab:results} summarizes the characteristic parameters derived from both the Radial and Radial–Poloidal models. Although both approaches produce a similar geometric orientation for the Flux Rope, the Radial–Poloidal model provides a substantially better fit to the internal magnetic field, suggesting that the additional terms lead to a more accurate and physically consistent representation of the structure.

\newcolumntype{Y}{>{\centering\arraybackslash}X}

\begin{table}[H]
    \centering
    \begin{tabularx}{\textwidth}{l Y Y}
        \toprule
        \textbf{Variable} & \textbf{Radial Model} & \textbf{Radial–Poloidal Model} \\ \midrule
        $\phi$ (Angle with x–z plane) & $-30.72^{\circ}$ & $-34.38^{\circ}$ \\
        $\theta$ (Angle with x–y plane) & $43.48^{\circ}$ & $37.52^{\circ}$ \\ 
        \midrule
        $a$ (Section Major Axis) [AU] & 0.1395 & 0.09077 \\
        $b$ (Section Minor Axis) [AU] & 0.08556 & 0.05564 \\
        $e$ (Elliptic section eccentricity, $a/b$) & 0.6133 & 0.6129 \\
        \midrule
        $R^{2}$ (Fitting Quality Factor) & 0.3782 & 0.6996 \\
        \bottomrule
    \end{tabularx}
    \caption{Characteristic parameters derived from the Radial and Radial–Poloidal models. The latter yields a substantially better fit quality, as indicated by the $R^{2}$ factor.}
    \label{tab:results}
\end{table}

From these results, we conclude that while both models capture the global geometry of the Flux Rope equally well, the Radial–Poloidal model achieves a significantly improved representation of the internal magnetic field, providing a more accurate and physically reliable description of the system.

\section{Conclusions}

In this work, we have presented a generalized elliptical–cylindrical flux–rope model that relaxes the purely radial assumptions imposed in previous approaches by allowing a nonzero poloidal contribution to the internal magnetic field. Together with this new formulation, we developed a two–step reconstruction algorithm that decouples the geometric configuration of the flux rope from the fitting of its internal magnetic structure.

This methodological innovation eliminates the degeneracy between orientation and physical parameters inherent in earlier global–fit techniques, improving numerical stability and enabling a systematic exploration of the parameter space. The analytical determination of the spacecraft trajectory within the flux rope, followed by a robust local optimization of the magnetic–field parameters, ensures convergence and physical consistency.

Application of the method to an in–situ event observed by PSP demonstrates that both the traditional radial model and the proposed radial–poloidal model reproduce the global geometry of the flux rope equally well. However, the radial–poloidal model achieves a substantially better fit to the internal magnetic field, as indicated by the higher $R^2$ values, suggesting that the additional poloidal terms provide a more accurate and physically realistic description of the system.

These results highlight the potential of the proposed methodology for large–scale studies of interplanetary coronal mass ejections (ICMEs), where robust and automated reconstruction techniques are essential for advancing our understanding of their magnetic structure and improving space–weather forecasting capabilities.

\bibliographystyle{spr-mp-sola}
\bibliography{references}  

\end{document}